\documentclass[11pt]{article}
\usepackage{fleqn,cospar}
\input{epsf}
\title{CURRENT STATUS OF COSMOLOGICAL MODELS WITH MIXED DARK MATTER}

\author{E.V. Mikheeva\address{Astro Space Center of Lebedev Physical 
Institute (ASC FIAN),\\ 
84/32 Profsoyuznaya st, Moscow, 117810, Russia}}

\begin{document}

\maketitle

\begin{abstract}
An analysis of cosmological mixed dark matter models in spatially 
flat Friedmann Universe with zero $\Lambda$-term is presented. 
We argue that the introduction of cosmic gravity waves helps to satisfy 
observational constraints. The number of parameters is equal to 5, they are 
(1) $\sigma_8$, the dispersion of the mass fluctuations in the
    sphere with radius $8\,h^{-1} Mpc$, 
(2) $n$, the slope of the density perturbation spectrum, 
(3) $\Omega_\nu$, the density of hot dark matter, 
(4) $\Omega_b$, the density of baryons, and 
(5) $h$, the Hubble constant $H_0=100\;h\;km\,s^{-1}Mpc^{-1}$.  
The cold dark matter density parameter is equal to $\Omega_{cdm} 
=1-\Omega_\nu -\Omega_b$. The analysis of models is based on the 
confrontation with the mass function of clusters of galaxies and 
the CMB anisotropy. The implication of Press-Schechter formalism 
allowed to constrain $\sigma_8=0.52 \pm 0.01$. 
This normalisation of the spectrum of density perturbations has been 
used to calculate numerically the value of the large scale CMB 
anisotropy and the relative contribution of cosmological 
gravitational waves, T/S. We found that increasing $\Omega_\nu$ 
weakens the requirements to the value of T/S, however even for 
$\Omega_\nu\le 0.4$ the models with $h+n\ge 1.5$ suggest 
considerable abundance of gravitational waves, T/S${}^>_\sim 0.3$. 
In models with $\Omega_\nu\le 0.4$ and scale-invariant spectrum of 
density perturbations ($n=1$), T/S${}^>_\sim 10(h-0.47)$. 
Minimisation of the value T/S is possible only in the range of 
the red spectra ($n<1$) and small $h$ ($<0.6$). However 
the parameter $\Omega_\nu$ is strongly constrained by $\Delta T/T$ 
data on the first acoustic peak of Sakharov oscillations. 
Assuming that T/S $\in[0,3]$ and taking into account observational 
data on the primordial nucleosynthesis and the amplitude of 
the first acoustic peak we constrain the model parameters. 
We show that the considered models admit both moderate red and 
blue spectra of density perturbations $n\in [0.9,1.2]$ with rather 
high abundance hot dark matter, $\Omega_\nu \in [0.2,0.4]$. 
Any condition, $n<0.9$ or $\Omega_\nu<0.2$, decreases the relative 
amplitude of the first acoustic peak for more than 30$\%$ in 
comparison with its height measured by BOOMERanG and MAXIMA-1.
\end{abstract}

%%%%%%%%%%%%%%%%%%%%%%%
\section*{INTRODUCTION}
%%%%%%%%%%%%%%%%%%%%%%%

A great success of observational cosmology of the last decade has allowed
to go from a purely theoretical investigation of possible cosmological 
models to testing them and constraining their parameters. Currently, 
the total number of model parameters exceeds ten, they can be separated 
in few groups:  
\begin {itemize}
\item 
The parameters related to the geometry of the Universe -- $\Omega_{total}$,
Hubble parameter $H_0$, and three-dimensional curvature $k$.
\item 
The parameters related to the nature and abundance of matter (including 
dark matter and baryons) in the Universe and the value of cosmological
constant ($\Lambda$-term), called often as "the dark energy". They are 
the matter density parameters in the Universe $\Omega_{matter}$ (the sum of 
$\Omega_{cdm}$, $\Omega_\nu$ and $\Omega_b$), the number of massive 
species of neutrinos $N_\nu$, the energy density of $\Lambda$-term 
$\Omega_\Lambda$, and other parameters of the dark energy
(e.g. quintessence, its equation of state $p=w\rho$, etc).
\item 
The group of parameters related to the inflationary stage of the Universe, 
which describes the power spectra of metric perturbations generated. In the 
simplest version of inflation based on the theory of scalar field these 
parameters are the amplitudes and the slopes of density perturbation and 
gravitational wave spectra.
\end{itemize}

A complete analysis of the cosmological model in constraining all mentioned 
above parameters (assumed initially as free ones) is currently a difficult 
task since it requires as excellent technical facilities as a special 
study of different sets of observational data for their completeness and 
independence. In view of the imperfection of currently available 
observational data one usually assumes a simplified model with restricted 
number of the parameters (and fixed others) and test it by a selected 
set of observational data.

The goal of the present analysis is to consider a matter-dominated model of 
the Universe and constrain its parameters by three fundamental tests which 
have different characteristic scales. These observational tests are the large 
scale CMB anisotropy ($\sim 1000\; h^{-1} Mpc$), the amplitude of the first 
acoustic peak of Sakharov oscillations ($\sim 100\; h^{-1} Mpc$) and the mass 
function of clusters of galaxies ($\sim 10\; h^{-1} Mpc$). {\em The aim of 
the analysis is to clarify the modern status of such a model: whether it can 
explain the observations or not, and thus, whether the $\Lambda$CDM is the only 
satisfactory model.}

To answer this question we considered a family of matter-dominated 
cosmological models with the following free parameters:
\begin{itemize}
\item $\sigma_{8}\in [0.47,0.61]$, (15 models with step 0.01);
\item $n\in [0.8,1.4]$, (7 models with step 0.1); 
\item $\Omega_\nu \in [0,0.4]$, (5 models with step 0.1);
\item $\Omega_b \in [0.01,0.11]$, (6 models with step 0.02);
\item $h \in [0.45,0.7]$, (6 models with step 0.05).
\end{itemize}
Altogether we had 18900 variants of the model. The broad intervals
for parameter variations are specially selected to find all possible 
solutions for the given tests.

The first parameter, $\sigma_8$, is a more accurate normalisation 
of density perturbation spectrum than COBE one, since a probable  
contribution of primordial gravitational waves into the COBE measured 
signal. The latter is the only available today data which we can use
to find a signature of cosmic gravitational waves.

The second parameter, $n$, generalises the scale invariant spectrum of 
density perturbations (Harrison-Zeldovich spectrum) in the simplest way. 
A power-law spectrum with $n<1$ (the "red" one) is predicted in a 
broad range of inflationary models (Linde, 1983, Lucchin and Matarrese, 
1985). In other models it can be both "blue" ($n>1$) and non-power-law 
(the noncomplete list of references includes: Carr and Gilbert (1994), 
Gottlober and M\"ucket (1993), Lesgourgues et al. (1997), Lukash and 
Mikheeva (2000), Lukash et al. (2000), Melchiorri et al. (1999), 
Mollerach et al. (1994), Semig and M\"uller (1996)). Here we consider
the power-law spectra with a free slope index $n$ within the range.

The introduction of hot dark matter is related with a recent discovery 
of the atmospheric neutrino oscillations, which demonstrates that 
at least one species of neutrino is massive. A careful theoretical 
analysis of all neutrino experiments argues for three massive species, 
but this result still requires a confirmation. In the numerical analysis 
we assumed that $N_\nu=1$. Actually, the value of $N_\nu$ becomes 
important only at small scales ($\,{}^<_\sim 1  \;h^{-1} Mpc$,  e.g. in
the analysis of $Ly_\alpha$-absorbers, the test which we do not consider 
here). 

The fourth parameter, $\Omega_b$, influences the transfer function 
slightly, so, neither mass function of clusters of galaxies nor large 
scale CMB anisotropy can reveal its value with a high confidence 
level. Hovewer, $\Omega_b$ is extremely important for the amplitude of 
the first acoustic peak of Sakharov oscillations.

The last parameter is Hubble constant. Till now there is some
discrepancy in its measurements (Saha et al., 1999, Jha et al., 1999). 
It is clear that $h$ cannot be smaller than 0.5, but on the other hand, 
the large value (${}^>_\sim 0.65$) will contradict the age of the oldest 
globular clusters.

In our analysis there are two derived parameters. The first of them is
the abundance of cold dark matter, $\Omega_{cdm} = 1-\Omega_\nu-\Omega_b$.
The second one is a relative contribution of primordial gravitational waves
into large scale CMB anisotropy, T/S. The latter parameter can be 
theoretically calculated in any inflationary model. T/S is not discriminated 
by inflation and vary in a broad interval (we assume T/S$\in[0,3]$ to take
into account all possible values). For the majority of inflationary models 
there is a simple relation between the spectrum slope of primordial 
gravitational waves, $n_T$, and T/S:
\begin{equation}
{\rm T}/{\rm S}\simeq -6n_T,
\end{equation}
where the numerical coefficient slightly depends on scale ( T/S is
calculated here for $10^o$ angular scale). Eq.(1) allows to replace one 
parameter (T/S) by another ($n_T$). It is interesting to note that in 
power-law inflation there exists an additional relation between $n$ and 
$n_T$, so in this model T/S can be expressed in terms of $n$. However, 
generally such relation is absent (see, e.g., Lukash et al. (2000), 
Lukash and Mikheeva(2000)).
 Thus, instead of considering additional 
constraints related with given inflationary model, we prefer to consider 
$n$ and T/S as independent variables within their ranges.

%%%%%%%%%%%%%%%%%%%%%%%%%%%%%%%%%%%%%%%%%%
\section*{MASS FUNCTION OF GALAXY CLUSTERS}
%%%%%%%%%%%%%%%%%%%%%%%%%%%%%%%%%%%%%%%%%%

\begin{figure}
\epsfxsize=85mm
\centerline{\epsfbox{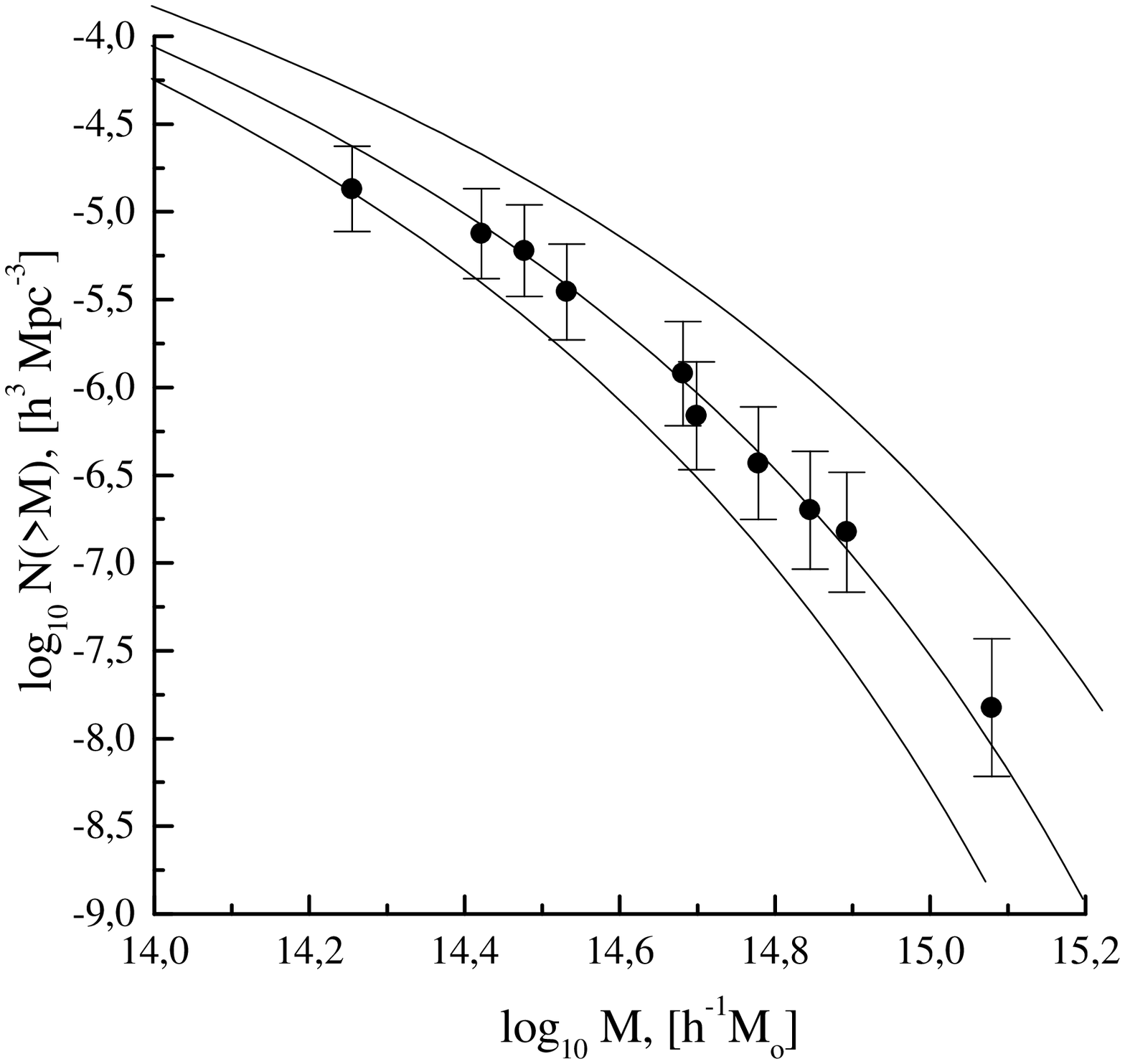}}
\begin{center}
\parbox{85mm}{\sf 
Fig. 1. $N(>M)$ as a function of $\sigma_8$ ($n=1$, $\Omega_\nu=0$, 
$\Omega_b=0.05$, $h=0.5$). The upper curve corresponds to the model with 
$\sigma_8=0.61$, the middle one -- $\sigma_8 = 0.52$, the lower curve -- 
$\sigma_8=0.47$.}
\end{center}
\end{figure} 

The number of massive halos with the mass larger than $M$ can be calculated 
with help of Press-Schechter (P\&S) formalism (Press and Schechter, 1974):
\begin{equation}
N\left(>M\right) = \int\limits_M^\infty\frac{dn}{dM^\prime}dM^\prime,
\end{equation}
where 
\begin{equation}
\frac{dn}{dM}=
\sqrt{\frac 2{\pi}}\frac{\rho_0\delta_c}{M}\frac 1{\sigma^2_R}\left
\vert\frac{d\sigma_R}{dM}\right\vert\exp\left(-\frac{\delta^2_c}{2
\sigma^2_R}\right),
\end{equation}
where
$M=4\rho_0R^3/3$, 
$\rho_0$ is the matter density in the Universe, 
$\delta_c$ is a critical density contrast necessary to form a halo 
($\delta_c=1.686$,
 see Gunn and Gott (1972)), 
$\sigma_R$ is a dispersion of linear density perturbations in the sphere
of radius $R$ which is an integral of the power spectrum of density 
perturbations
\begin{equation}
\sigma_R^2 =\int_0^\infty \Delta_k^2W^2(kR)\frac{dk}{k},
\end{equation}
where $\Delta^2_k = \frac 1{2\pi^2}P(k)T^2(k)$ is a dimensionless 
power spectrum, 
$P(k) = Ak^{n+3}$ -- a primordial spectrum of density perturbations, 
$A$ -- a normalisation constant,
$T(k)$ -- a transfer function which depends on $\Omega_\nu$, $\Omega_b$ and 
$h$ (we used the analytic approximation of the transfer function given by 
Novosyadlyj et al. (1999)), 
$W(kR)$ -- Fourrier transform of {\it top-hat} filter, 
$W(x) = 3\left(\sin x-x\cos x\right)/x^3$. 

\begin{figure}
\epsfxsize=85mm
\centerline{\epsfbox{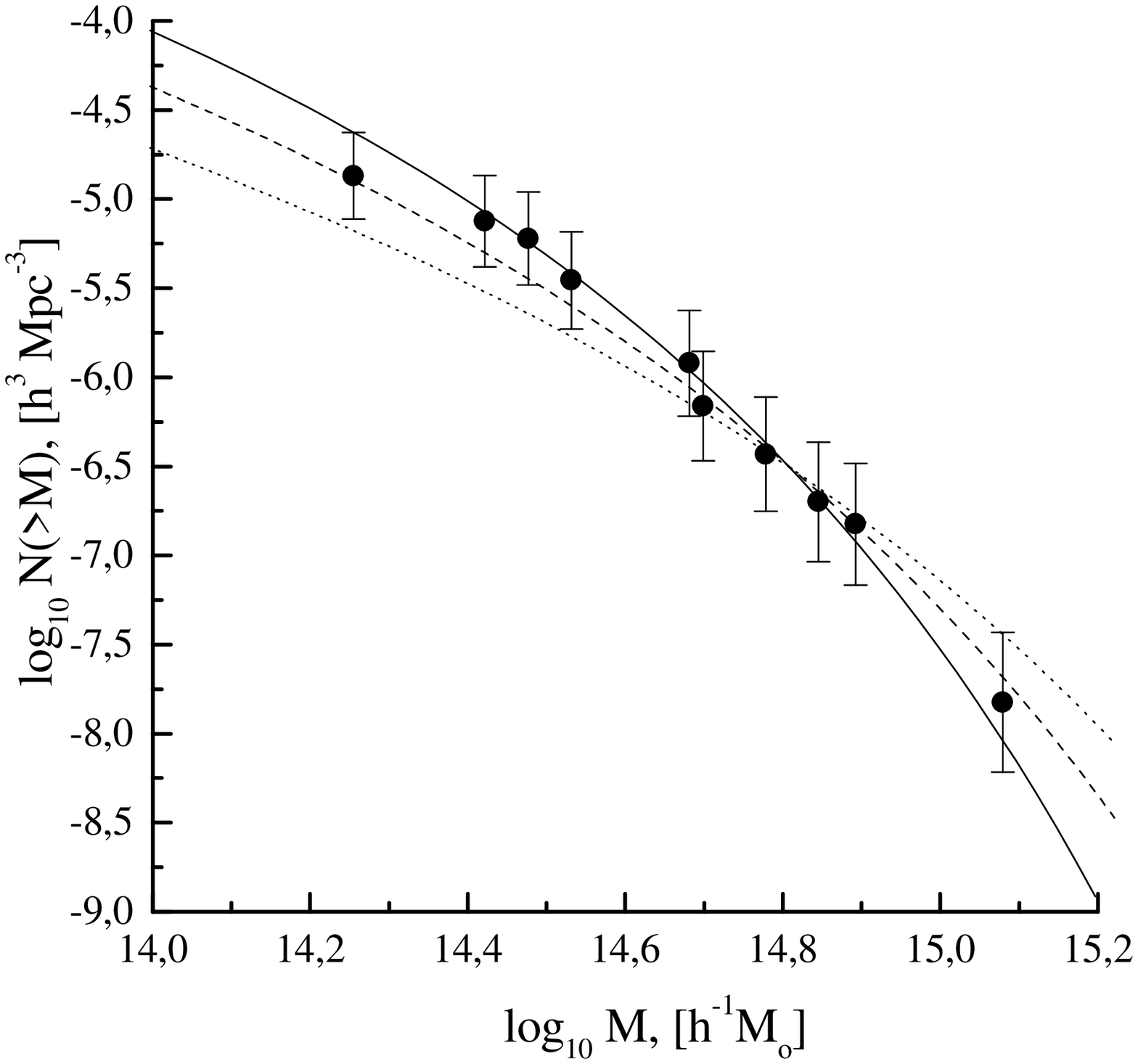}}
\begin{center}
\parbox{85mm}{\sf 
Fig. 2. $N(>M)$ as a function of $\Omega_\nu$ ($\sigma_8=0.52$, $n=1$, 
$\Omega_b=0.05$, $h=0.5$). The solid line corresponds to the model with 
$\Omega_\nu=0$, the dashed one -- $\Omega_\nu = 0.2$, the lower line -- 
$\Omega_\nu=0.4$.}
\end{center}
\end{figure}

The Figure 1 illustrates how the function $N(>M)$ depends on $\sigma_8$ 
with other parameters fixed. As it was mentioned in the Introduction, 
$\sigma_8$ affects efficiently  the height of P\&S curves which one can use
to determine $\sigma_8$. The Figure 2 demonstrates the behaviour of $N(>M)$ 
as a function of $\Omega_\nu$. Taking a given $\sigma_8$ (0.52 in the 
Figure 2) fixes the amplitude of P\&S curve at the mass value in the sphere 
with radius $8\; h^{-1} Mpc$, while different values of $\Omega_\nu$ change 
the slope of the curve. In  both Figures the circles with errorbars mark the 
observational data. For the rest parameters, $N(>M)$ varies with a much smaller 
amplitude and we do not adduce the corresponding figures. The observational 
data in the Figures 1 and 2 are taken from Bahcall and Cen (1993). 

To analyse the cosmological models we have used the standard $\chi^2$ method.
At the first step we have calculated $\chi^2$ for each model (this function
is $\chi^2$ distributed with 5 degrees of freedom; 
$N_{d.f.} = N_{obs.points} - N_{parameters} = 10-5$). 
It turns out that the model with the minimal $\chi^2$ ($\chi^2_{min}=1.43$) 
has the following parameters: $\sigma_8=0.52$, $n=1.3$, $\Omega_\nu=0.3$, 
$\Omega_b=0.01$, $h=0.7$.
To evaluate the confidence level of this result we took into account 
the mathematical theorems available for $\chi^2$ and constructed other 
distributions for each parameter 
$\Delta\chi_1^2(a) = \chi^2_a - \chi^2_{min}$, where $a$ is one of the five
cosmological parameters, $\chi^2_a$ is a minimal $\chi^2$ on the hypersurface 
$a=const$ in 5-dimensional model space. $\Delta\chi_1^2(a)$ is also $\chi^2$
distributed but has the only degree of freedom. The confidence levels
1, 2 and $3\sigma$ correspond to $\Delta\chi_1^2(a)=1$, 4, and 9
respectively. $\Delta\chi_1^2(a)$ for $a=\sigma_8$ is shown in the Figure 3
(the curve marked by $"\times"$).
One can see that $\sigma_8 = 0.52 \pm 0.01$. The uncertainties of the 
Press-Schechter formalism and data systematics enhances the total errorbar 
to 0.04 (Eke et al., 1996, Borgani et al., 1999). 

To evaluate the accuracy of the approximation of the transfer function
and its influence on the output we have repeated this statistical analysis 
(see Figure 3, the line marked by $"+"$) with another analytic expression 
for the transfer function (Eisenstein and Hu, 1999) and found that both 
approximations lead to the same result. 

As to the other cosmological parameters ($n$, $\Omega_\nu$, $\Omega_b$, 
$h$) their $\Delta\chi_1^2(a)$ are under $1\sigma$ confidence level, 
therefore they cannot be constrained by the cluster data only. 
{\em So, as a result of this part of our analysis we have obtained 
the high 
accuracy normalisation of the density perturbation spectra}. 
It is evident that the obtained value of $\sigma_8$ depends on 
observational data used and can be slightly different if one uses other
data (e.g. Borgani et., 1999). 

\begin{figure}
\epsfxsize=85mm
\centerline{\epsfbox{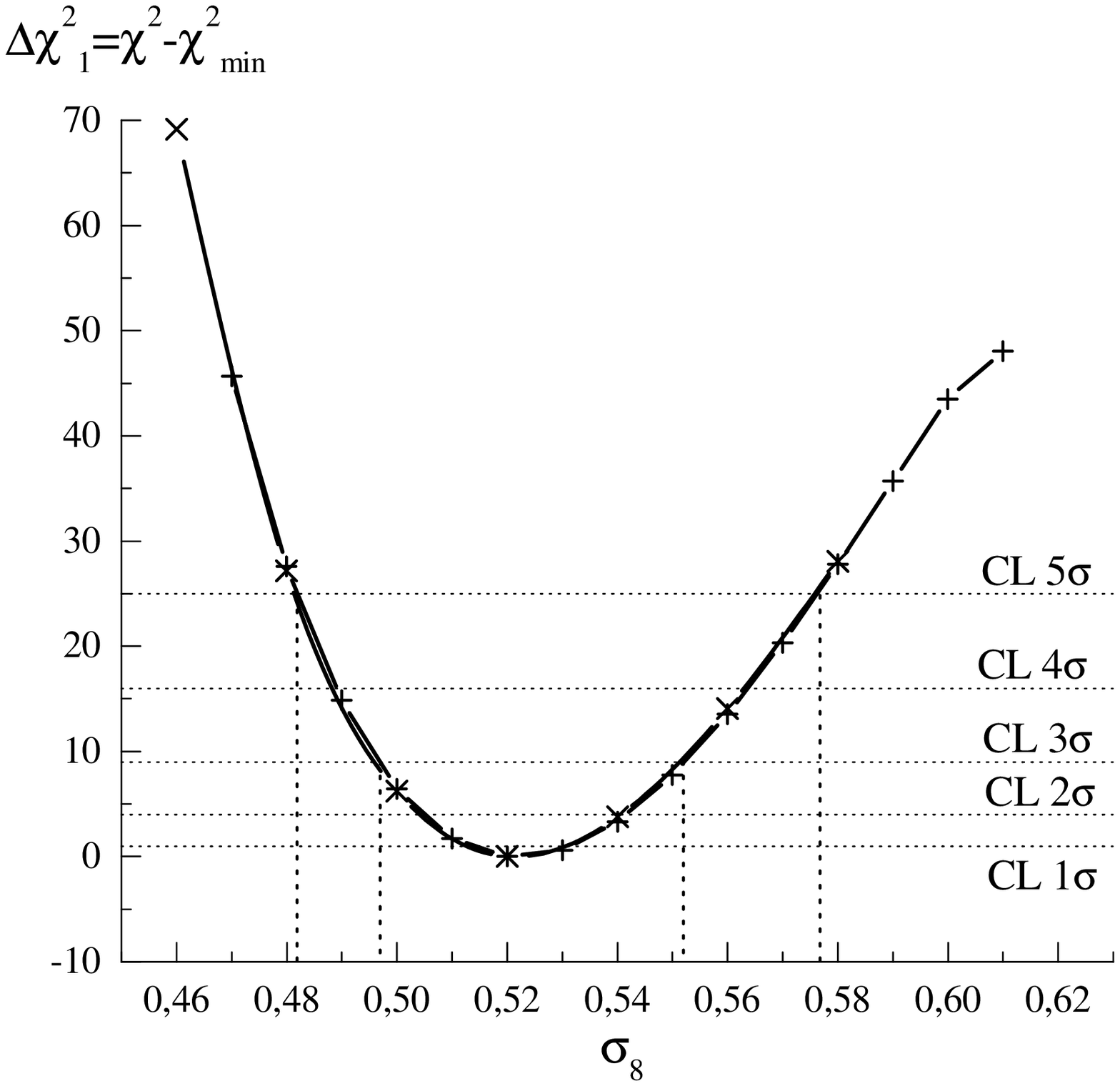}}
\begin{center}
\parbox{85mm}{\sf 
Fig. 3. $\Delta\chi^2_1\equiv \chi^2-\chi^2_{min}$ as a function of
$\sigma_8$.
 The curve marked by "+" corresponds to the calculations with the transfer 
 function of Eisenstein and Hu (1999). The curve marked by "$\times$" 
 corresponds to the transfer function from Novosyadlyj at al. (1999). 
The horizontal lines mark the different confidence levels.}
\end{center}
\end{figure}

%%%%%%%%%%%%%%%%%%%%%%%%%
\section*{CMB ANISOTROPY}
%%%%%%%%%%%%%%%%%%%%%%%%%%%%%%%%%%%%%%%%

\subsection*{Large Scale CMB Anisotropy}
%%%%%%%%%%%%%%%%%%%%%%%%%%%%%%%%%%%%%%%%

The large scale CMB anisotropy is related with metric perturbations
(Sachs and Wolfe, 1967):
\begin{equation}
\frac{\Delta T}T\left(\vec e\right)=
\frac 12\int_R^E\frac{\partial h_{ik}}{\partial\eta}e^ie^kd\eta,
\end{equation}
where $E$ and $R$ are emission and reception moments of time, $h_{ik}$ is a
metric perturbation tensor, $\partial/\partial\eta$ is the conformal 
time derivative, the integral goes along the null geodesics.

The relative contribution of cosmic gravitational waves into the large 
scale CMB 
anisotropy (T/S) can estimated as follows:
\begin{equation}
\left\langle\left(\frac{\Delta T}{T}\right)^2\right\rangle_{10^o}={\rm S}+
{\rm T}= {\rm S}\left(1+\frac{\rm T}{\rm S}\right)\simeq 1.1\times 10^{-10}, 
\end{equation} 
where S is the contribution of the density perturbations:
\begin{equation}
S=\sum_{\ell=2}^{\infty}S_{\ell}W_{\ell},
\end{equation}
where
\begin{equation}
S_{\ell}=\frac{2\ell+1}{64\pi}AH_0^{n+3}
\frac{\Gamma(3-n)\Gamma(\ell+(n-1)/2)}{\Gamma^2(2-n/2)\Gamma(\ell+(5-n)/2)},
\end{equation}
and
\begin{equation}
W_{\ell}=\exp\left[-\left(\frac{2\ell+1}{27}\right)^2\right].
\end{equation}
We evaluate the normalisation constant $A$ from Eq.(4) for $\sigma_8$, 
$W_{\ell}$ is COBE DMR window function. 

Figure 4 
illustrates T/S as a function of $n$ and $\Omega_\nu$ for $\sigma_8=0.52$,
$\Omega_b=0.05$ and $h=0.5$. The thick line represents the T/S-$n$ relation
for the power-law inflation. It can be seen that small values
of T/S favour red spectra of density perturbations.

\begin{figure}
\epsfxsize=85mm
\centerline{\epsfbox{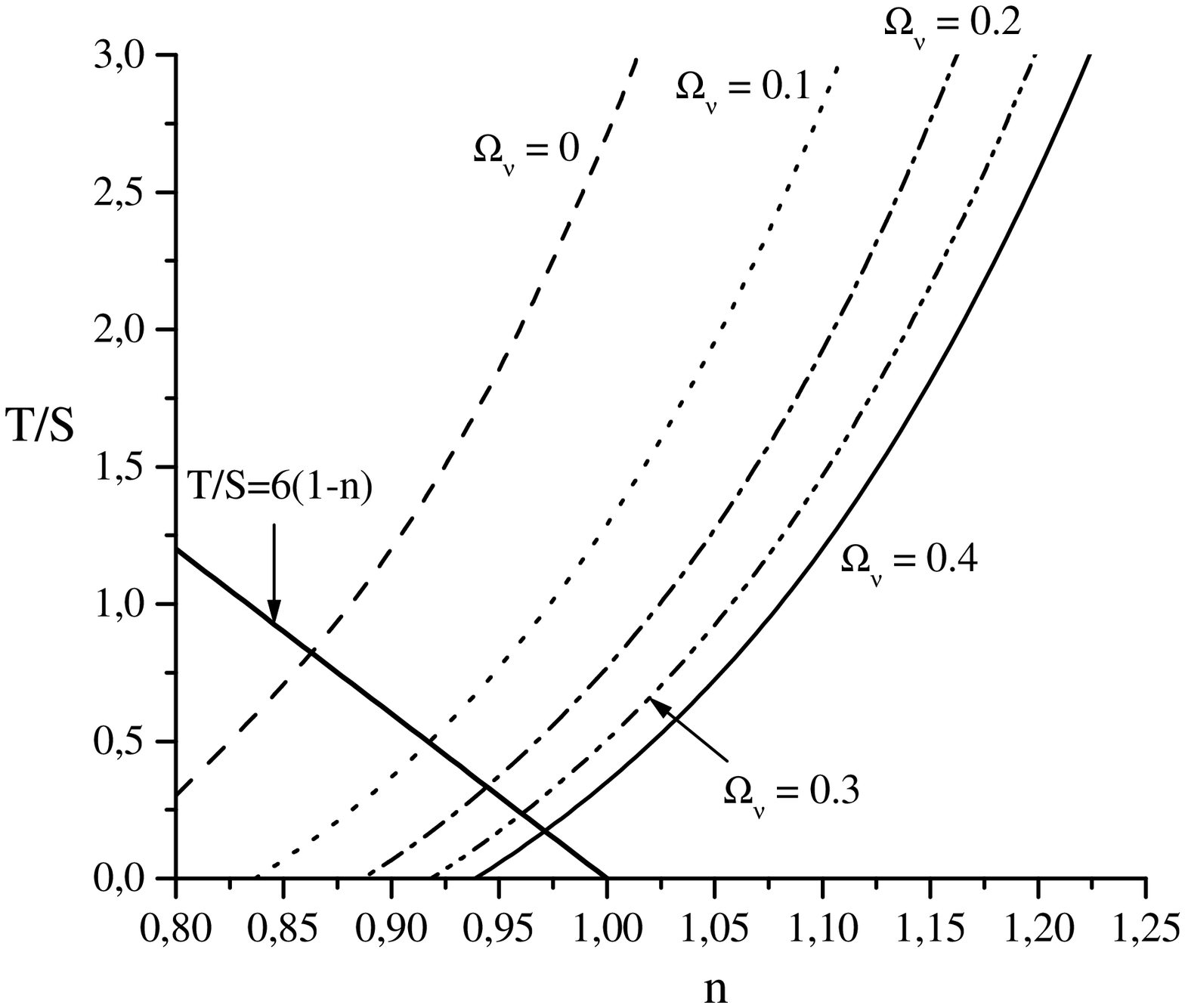}}
\begin{center}
\parbox{85mm}{
\sf Fig. 4. T/S as a function of $n$ and $\Omega_\nu$ ($\sigma_8=0.52$, 
$\Omega_b=0.05$, $h=0.5$).}
\end{center}
\end{figure} 

The Figure 5 demonstrates T/S as a function of $n$ and $h$ for 
$\sigma_8=0.52$, $\Omega_b=0.05$, and large $\Omega_\nu=0.4$.
The value T/S increases linearly with $h$ and decreases with $\Omega_\nu$ 
growing, therefore the curves T/S with the maximum parameter $\Omega_\nu =
0.4$ can be used to put the lower limit on T/S (see Conclusions).

Taking a moderate T/S$<0.5$ and nearly flat power spectrum ($0.92\le n\le 
1.02$), we put an upper limit on the Hubble constant, $h<0.6$, and lower limit
on the hot dark matter abundance, $\Omega_\nu >0.1$. However, the hardest 
constraint for the parameter $\Omega_\nu$ can be got when we confront 
the amplitude of the first acoustic peak in $\Delta T/T$ ($\ell\simeq 200$) 
with the observational data.    

\begin{figure}
\epsfxsize=85 mm
\centerline{\epsfbox{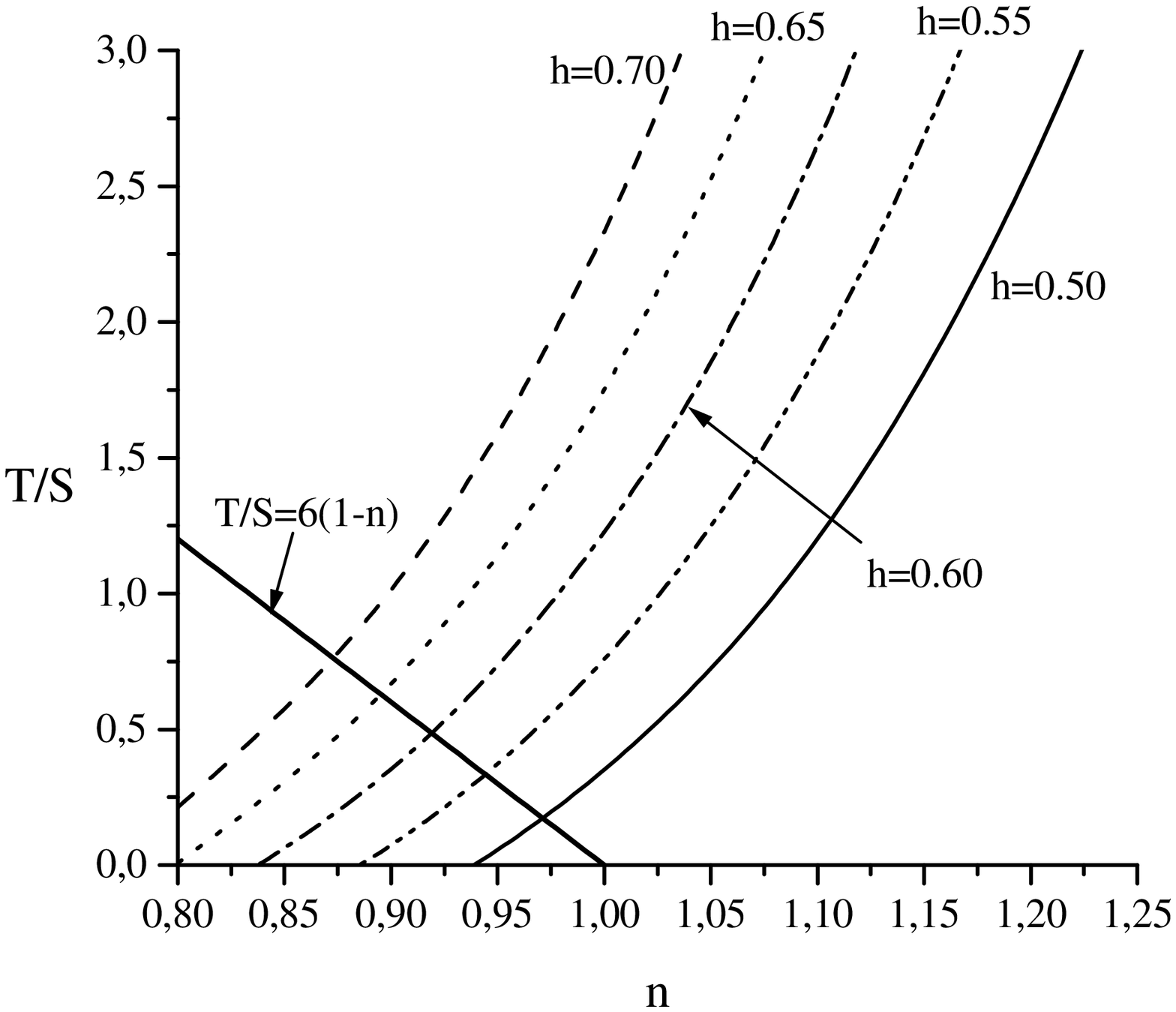}}
\begin{center}
\parbox{85mm}{
\sf Fig. 5. T/S as a function of $n$ and $h$ ($\sigma_8=0.52$, 
$\Omega_\nu=0.4$, $\Omega_b=0.05$).}
\end{center}
\end{figure}

For $\sigma_8=0.52$, $\Omega_b=0.05$ and $h=0.5$ we propose the following 
approximation for T/S (the accuracy is better than 11 \% for 
$0.1\le$T/S$\le 3$):
\begin{equation}
{\rm T}/{\rm S}=\frac{30(n-0.7)^2}{10\Omega_\nu+1}+10\Omega_\nu\left(
n^{3/2}-1.06\right).
\end{equation}

%%%%%%%%%%%%%%%%%%%%%%%%%%%%%%%%%%%%%%%%%%%%
\subsection*{The Medium Scale CMB Anisotropy}
%%%%%%%%%%%%%%%%%%%%%%%%%%%%%%%%%%%%%%%%%%%%
To compare the height of the first acoustic peak generated in our models 
with its value measured by BOOMERanG (Lange et al., 2000) and MAXIMA-1 
(Balbi et al., 2000) we introduce the
 dimensionless parameter $\Re$ which 
is the amplitude of the peak in comparison with
 the large scale CMB anisotropy
amplitude: 
$\Re \equiv \Re_{\ell=200} / 1.1\times 10^{-10}$, where
$\Re_\ell\equiv\ell (\ell+1)S_\ell /(\ell+0.5)$; $\Re\simeq 5$ for 
BOOMERanG.

Evidently, $\Re$ decreases with T/S growing (and other parameters fixed).
E.g. for normalised by $\sigma_8$ standard Cold Dark Matter models 
($\Omega_\nu=0$, $n=1$, $\Omega_b=0.08$, $h=0.5$) the relative amplitude 
of the acoustic peak decays by a factor T/S$+1\simeq 4$ and practically 
disappears.

The more efficient ways to enhance the first acoustic peak is a transition 
to the red power spectra and/or high abundance of the hot matter. The role of
the blue spectra becomes important when the parameter $\Omega_\nu$ rises
up (since the red spectra will contradict to the condition T/S$\ge 0$).
The results of the derivation are presented in the Figure 6. The condition
for a "considerable" acoustic peak ($\Re\ge 3.5$) with the instantaneous 
Big Bang nucleosynthesis constraint for the baryonic density, leaves us 
with the standard family of
 the power spectra ($n\in [0.9,1.2]$) but 
requires high fraction of the
 hot matter ($\Omega_\nu \in [0.2,0.4]$) in the 
class of the models considered.

Varying $\Omega_b$ we found that increasing $\Omega_b$ by factor of two
results in a significant rise of the acoustic peak independently on $n$ 
and $\Omega_\nu$, so for models with $\Omega_b=0.1$, $n\in[0.9,1.2]$ and 
$\Omega_\nu\in[0.2,0.4]$: $4\le\Re\le 5.5$. As to the parameter $h$, 
its increasing decreases the acoustic peak amplitude. Summarizing, 
we can conclude that MDM models with gravitational waves can marginally 
reproduce the BOOMERanG result indeed.      

%%%%%%%%%%%%%%%%%%%%%%
\section*{CONCLUSIONS}
%%%%%%%%%%%%%%%%%%%%%%

\begin{figure}
\epsfxsize=85mm
\centerline{\epsfbox{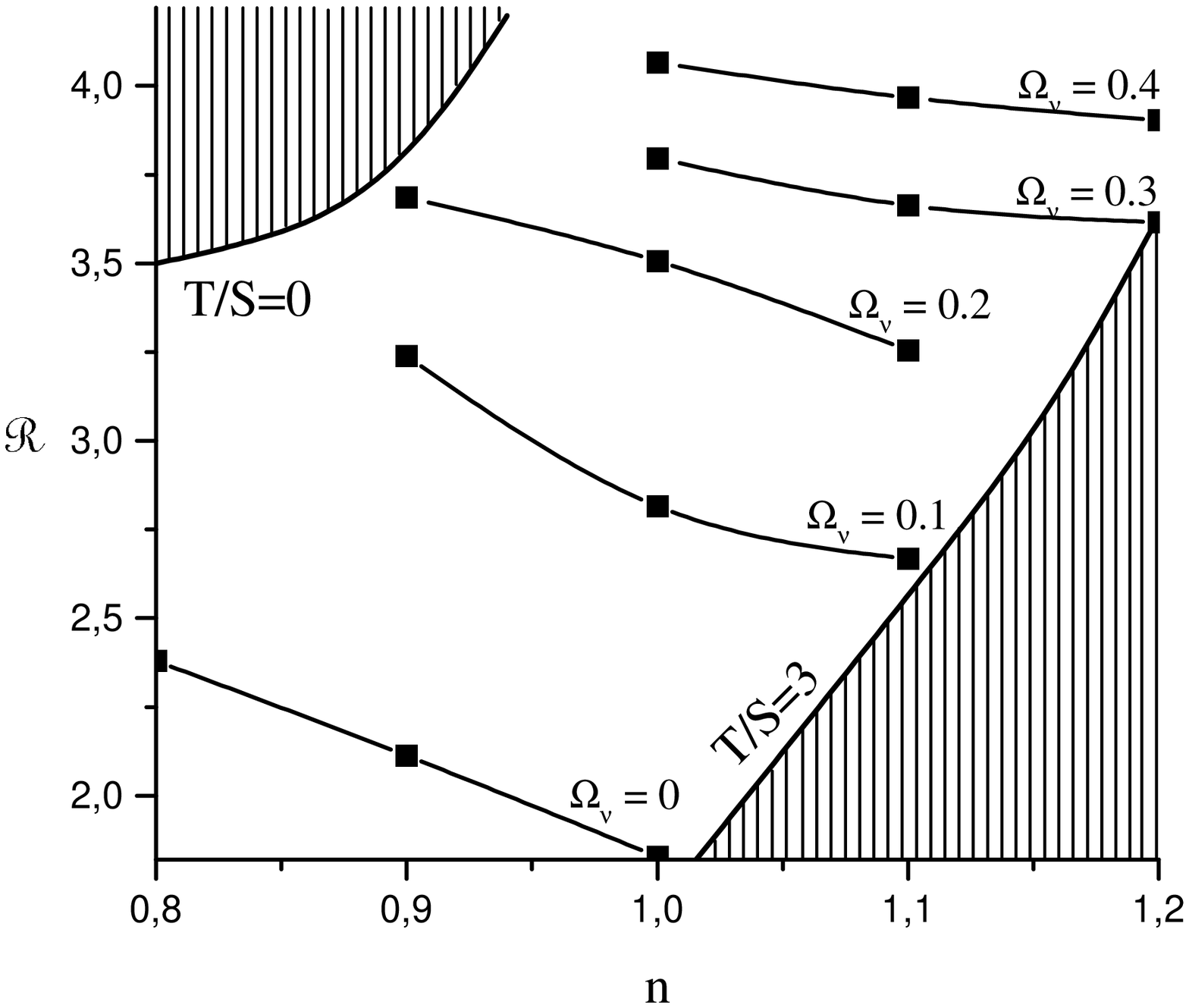}}
\begin{center}
\parbox{85mm}{
\sf Fig. 6. $\Re$ as a function of $n$ for different values of $\Omega_\nu$ 
($\Omega_b=0.05$, $h=0.5$, $\sigma_8=0.52$). Models in the non-shaded 
region have T/S$\in[0,3]$.}
\end{center}
\end{figure}

\begin{itemize}
\item The data on the galaxy cluster abundance determine the value $\sigma_8$ 
with a high accuracy, the other parameters ($n, \Omega_\nu, \Omega_b, h$) 
remain free within their ranges.    
\item None of the MDM models with $n=1$ and $T/S=0$ satisfies both 
normalisations, 
on the galaxy cluster abundance and large scale $\Delta T/T$
anisotropy, which leads either to rejection from the flat spectrum or to
the introduction of a non-zero T/S (or both).
\item Small values of T/S are realised for the red spectra ($n<1$) and 
moderate $h(<0.6)$, the violation of these conditions leads to a high 
T/S(${}^>_\sim 1$).
\item 
Increasing $\Omega_\nu$ weakens the requirement to the value of T/S, however 
even for $\Omega_\nu\le 0.4$ the models with $h+n\ge 1.5$ suggest considerable
abundance of gravitational waves: ${\rm T/S}^>_\sim 0.3$.
\item 
In models with $\Omega_\nu\le 0.4$ and scale-invariant spectrum of density 
perturbations ($n=1$): T/S${}^>_\sim 10(h-0.47)$.
\item 
In models with $\Omega_b=0.05$ and $h=0.5$ we have the approximation
formula for the primordial gravitational wave abundance (see Eq.(10)). 
\item 
In double-normalised models with T/S$>$0 the height of the acoustic peak is 
less than its `standard' value ($\Re=5$). The decrease of the parameter 
$\Re$ does not exceed 30\% in models with large $\Omega_\nu$ ($\in [0.2,0.4]$) 
and any spectrum slope, $n\in [0.9,1.2]$. Any condition, $n<0.9$ 
or $\Omega_\nu<0.2$, decreases the relative amplitude of the first 
acoustic peak for more than 30$\%$ (i.e. $\Re<3.5$ in models with
$\Omega_b=0.05$, $h=0.5$). The acoustic peak practically disappears
in CDM models.
\item
When growing the baryonic abundance the factor $\Re$ increases. 
The amplitude of the acoustic peak coincides with 
its 'standard' value ($\Re=5$) up to accuracy 10\% in models with 
$\Omega_b=0.1$ and both, blue spectra $n\in[1,1.2]$ and $\Omega_\nu\ge 0.3$ 
as well as moderate red spectra $n\in[0.9,1]$ and $\Omega_\nu\ge 0.2$.
\item
Thus, rising the parameter $\Omega_\nu$ up to the values in the interval 
$[0.2,0.4]$ solves the problem of the first acoustic peak in $\Delta T/T$, 
leaving the baryonic density within the primordial nucleosynthesis 
constraints.   
\end{itemize}

%%%%%%%%%%%%%%%%%%%%%%%%%%%
\section*{ACKNOWLEDGEMENTS}
%%%%%%%%%%%%%%%%%%%%%%%%%%%

The work was partly supported INTAS (grant 97-1192). 
The author is grateful to the Organizing Committee for hospitality. 

%%%%%%%%%%%%%%%%%%%%%%%%%%

\end{document}